# *Logic Programming as Scripting Language for Bots in Computer Games – Research Overview*

Grzegorz Jaśkiewicz

*Warsaw University of Technology, Poland*
(*e-mail:* `grzegorz@jaskiewi.cz`)

## Abstract

This publication is to present a summary of research (referred as $\kappa$-Labs[1]) carried out in author's Ph.D studies on topic of application of Logic Programming as scripting language for virtual character behavior control in First Person Shooter (FPS) games.

The research goal is to apply reasoning and knowledge representation techniques to create character behavior, which results in increased players' engagement.

An extended abstract / full version of a paper accepted to be presented at the Doctoral Consortium of the 30th International Conference on Logic Programming (ICLP 2014), July 19-22, Vienna, Austria

## Contents

---

[1] website available at `http://www.kappalabs.org`

## 1 Introduction

The First Person Shooter (FPS) is a type of video game where a player sees the level from the eyes of the character being played. *Counter-Strike* is one of the most popular and successful FPS games in the world. It is one of the top 10 online games on the Steam network with a daily peak of online players near 40.000[2] and it is played professionally around the world in e-sport leagues (Jana et al. 2007). Sample screenshots of the gameplay is presented in the fig. 1.

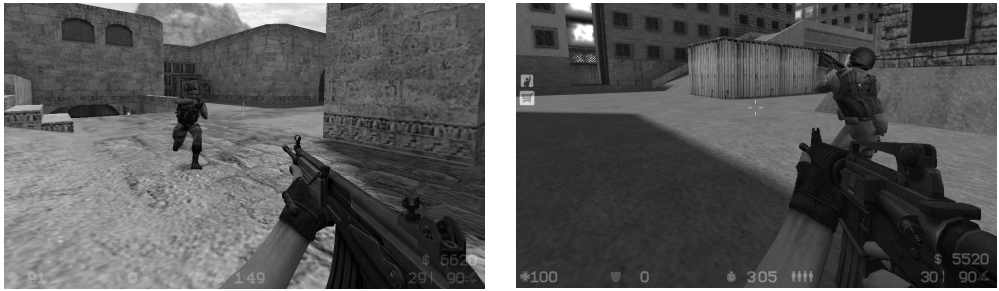

Fig. 1. First person perspective view in *Counter-Strike* game.

The game is about fight of terrorist (Ts) and counter-terrorist (CTs) forces. Players are divided into those two teams and try to fulfill objectives, which are dependent on a map type. Most commonly played map types are defusion (DE) and counter-strike (CS). On CS maps CTs has to rescue hostages, which are guarded by Ts. Example of such map is `cs_assault`, where hostages are held in a warehouse, or `cs_747`, where hostages are held in a Boeing plane on an airfield. In DE maps CTs has to prevent Ts from bombing some important location.

The game was created in 1994 and its first versions were entirely multiplayer[3]. The lack of ability to play in single player mode resulted in community-developed Artificial Intelligence (AI) algorithms for replacing human controlled opponents. The software which controls characters in a multiplayer game is usually called "bot".

*Counter-Strike* has large community of people who create modifications for the game (Kücklich 2005). As a result there are available many different bot implementations. Many of them are distributed as an open-source software. The research described in this article is based on the open-source bot implementation called *E[POD]*. The bot's source code was altered to allow scripting the bot behavior using Logic Programming. The name $\kappa Bot$ is used to refer to the bot which is an object of the research.

The idea of embedding a scripting language into video game is well-known practice in the game development industry. It allows to clearly divide responsibilities in project of creating a video game – game engine programmers and game world designers. In some cases game engine is not developed. Instead, the license is bought for usage in the particular product. The ability to script some aspects of gameplay makes such engine more flexible. Commonly, scripts are used to describe player interactions with game world, but character behavior is also part of this domain, hence there is a perfectly valid use-case for existence scripting tool for their behavior.

---

[2] source: daily statistics for *Counter-Strike* 1.6 provided by the Steam – online gaming platform, see `http://store.steampowered.com/stats/`, access date: 07.2013.
[3] game mode where multiple players participate in same game session; often played on a network (LAN or Internet).

`Lua`, `Python` and `Unreal Script` dominate as scripting languages for game engines (Anderson 2011). All of them are multiple paradigm languages: imperative, object-oriented, functional, procedural. However, none of them does allow logic programming by its syntax. This is distinct paradigm, which relies strongly on notion of reasoning and knowledge representation – the main concepts of rational agents. Virtual characters in FPS games could be regarded as rational agents, so the scripting language for character behavior in *κBot* allows logic programming.

## 2 Related works

### 2.1 Rule-based systems in video games development

The similarities presented in this section are based on application of rule-based systems for expressing a character behavior rules in video games. The most prominent examples of such solutions are:

- Generic Robot Language (Horswill 2000) – a simple functional language to define decision rules, which are compiled to `C++` code;
- Goal Oriented Action Planning (Orkin 2005) – a performance-oriented planning algorithm based on $A^\star$ algorithm. Uses pre- and post-conditions based on a predicate calculus for actions;
- Adaptive game AI with dynamic scripting (Spronck et al. 2006) – a machine learning algorithm used to optimize a selection of preprogrammed decision rules for fighting characters;
- Cognitive Modeling Language (Funge 1998) – a language integrated with planning using $A^\star$ algorithm. It allows specifying constraints through logical rules on solutions being searched;
- Avatar Definition Language (Anderson 2005) – an imperative language to script conditions for triggering state changes for an underlying finite state machine. The script represents a high-level logic of a bot functioning, while states are executing actions for achieving simple goals.

### 2.2 Frameworks for developing rational agents

Multiagent scientific field is well-developed. There exists many tools and methods, which facilitate software development in this paradigm.

The examples of platforms for general multiagent development are JADE (Bellifemine et al. 2005) and SOAR (Laird et al. 1987). Those could be possibly used for controlling virtual characters in a game environment (Laird 2001).

There is also a software platform dedicated for that purpose: Pogamut (Gemrot et al. 2009).

The examples of planning algorithms for rational agents which are based on knowledge representation and reasoning are:

- GOAL (Hindriks 2009) - agents derive their choice of action from their beliefs and goals,
- FLUX (Thielscher 2005) - programming framework for rational agents based on the fluent calculus (Thielscher 1998).

The variety of tools for developing rational agent is a motivation to apply some of them for characters in video games.

## 3 Research Overview

The goal of the research is to explore capabilities of logic programming as scripting language for controlling virtual characters in FPS video games. The reason for starting this research was a presence of logic programming in a multiagent programming domain.

### 3.1 Research status summary

We have choosen `Prolog` as scripting language for κ*Bot*, because it is a general-purpose, Turing-complete scripting language. This is important, because it gives programming flexibility, so that a programmer can construct arbitrary algorithms using `Prolog` script. Still, the κ*Bot* scripts can benefit from declarative syntax, which is property of `Prolog`.

We have chosen a SWI-Prolog (Wielemaker 2003) as `Prolog` interpreter, because of unrestrictive license and ease of embedding into software written in `C++`. The interpreter is not only a logic module, which computes a model for given rules and facts in order to make a decision. It is capable of invoking functions from the bot's code. We made such decision to enhance `Prolog` usage as programming language, not only as an inference method. Typically, some parts of the script define agent's reasoning rules and some are part of framework for reasoning, *i.e.* we use some generic predicates to divide inference rules into different generality levels: game rules, map type rules and map-specific rules.

We have used source code of *E[POD]* bot to create κ*Bot* bot. The original version of *E[POD]* relies on Brooks architecture (Zubek 2001) – the logic is divided into two layers with different level of abstraction:

- low-level reasoning – for expressing simple actions like moving around the map, attacking enemy, planting bombs, *etc*. Low-level reasoning algorithm tailored for particular task (*e.g.* planting a bomb) is called high-level action.
- high-level reasoning – for building bot behaviors by selecting of high-level actions.

The practice of separating task into several layers of abstraction is common practice in multiagent programming. For game bots two layers of abstraction is reasonable amount. However, for more complicated agents, more layers could be defined, *e.g.* robots, which must control their servomotors, may need additional hardware abstraction layers (Minsky 1986).

The goal is to develop a scripting module for the high-level reasoning layer. In first versions of κ*Bot*, *E[POD]*'s high-level decision-making algorithm was mimicked entirely by `Prolog` script. This was achieved by rewriting in `Prolog` all the rules and conditions which trigger change of high-level actions. Afterwards, rules for controlling bot behavior were divided into packages with different level of generality, *i.e.* general rules for playing the game, rules for playing on specific map type and rules for playing on concrete instance of the map. In recent versions κ*Bot* bot has predefined set of tactics scripted per map. Those tactics describe how to exploit map-specific features for gaining tactical advantage. For some maps they are even simulated bot negotiation for selecting collective team tactics, *e.g.* bot can vote according to its own, individual preferences, which tactic should it commit to.

### 3.2 Bot architecture elaborated during research

In this section we will show a core concept of the bot's architecture, which is scriptable in `Prolog`. Many specific details like bots' communication or collaboration are not covered by this section. However, any of those details are based upon concepts presented in this chapter.

Obviously, to make any decisions for any character the bot must learn information about the game environment it operates within. Relevant information are feed into `Prolog` script by invoking native predicates. Those can retrieve and transfer information, which are managed by the game engine, into `Prolog` interpreter, *e.g.* information about visible enemies, hearing footsteps, amount of ammunition and money available *etc.* An example of the declaration of such predicate is shown in the listing 1.

Listing 1. An example of native predicate declaration.
```
// predicate bot_in_fov(?botID, ?botID)
// checks if a character is in a field of view
// of another character. May have multiple goals.
PREDICATE_NONDET(bot_in_fov, 2) { /* ... */ }
```

Another source of information is a database of dynamic clauses, which is implemented with help of `assert` / `retract` predicates. Those information are used to implement the bot's memory. In contrary to information provided by the native predicates, memorized information can be changed freely by `Prolog` script. The example of information kept in bot's memory are, *e.g.*

- information about a map topology, which change infrequently, *e.g.* hiding spots, ambush points,
- information about past actions, *e.g.* bot should not try to buy weapons more then once a round,

High-level actions are started by invoking a native predicate in the script. The backtracking of a proof does not cancel once created action. In order to keep control over the script execution, the bot script should be constructed in such way that these action predicates will not be backtracked. This often leads to decision rules in following form:

Listing 2. An example of starting an action from script.
```
do_reasoning(BotID) :-
   should_do_action(BotId, Args_for_action),
   do_action(BotId, Args_for_action).
```

where `do_action` in listing 2 is a native predicate for starting some action. Term `should_do_action` is condition for this action; backtracking can occur while proving it. The conditional expression also provides arguments, which are passed to the action being invoked.

The high-level actions has two important features: motivations and continuations. A motivation is a logical condition expressed through `Prolog` terms. The action is executed as long as motivation is evaluated to truth value (`YES`) according to information, which the bot receives from the environment. Introducing the motivation improves control over the bot by `Prolog` scripts, because it is possible to interrupt execution of high-level action. This technique also improves the separation between bot's `C++` code and `Prolog` scripts, because script creator does not have to know detailed specification of any particular action.

An example of a motivation:

Listing 3. An example of motivation for action.
```
action_kill(
  BotID,
  EnemyID,
  and(bot_alive(EnemyID), danger_low(BotID))).
```

The action in listing 3 will make bot with identifier `BotID` try killing the bot with identifier `EnemyID` as long as invocation

```
call(and(bot_alive(EnemyID), not_in_danger(BotID))).
```

evaluates to YES. This condition expresses conjunction of two sub-subconditions: the character referenced by `EnemyID` is alive and the character referenced `BotID` is not in danger. At the moment of checking the motivation both variables are instantiated — `BotID` is owner of task and `EnemyID` is the target to eliminate.

Continuations are simple form of planning and defining complex behaviors consisting of sequence of several high-level actions. There could be assigned a continuation to any high-level action. A continuation is a term in `Prolog` script which is executed after high-level action is successfully completed. As a result of executing continuation there could be created a new action to be executed. The new action could have its own motivation and continuation. Example of continuation usage:

Listing 4. An example of continuation for action.
```
action_goto(BotID, Wp, andThen(
  action_liberate_hostages(BotID)).
```

The action in listing 4 will create an action for the character with identifier `BotID`, which will cause him go to a waypoint with identifier `Wp`. After reaching waypoint the continuation is started by executing

```
call(action_liberate_hostages(BotID)).
```

This will create new action for the character to liberate hostages.

The schematics of dependencies between system components has been presented in diagram 2.

The dependencies in diagram 2 are following:

1. high-level actions change environment (*e.g.* move an agent),
2. high-level actions read environment state,
3. script execution may create new high-level actions,
4. motivations may cease action execution; continuations may create new actions,
5. reasoning, motivations and continuations may read and change contents of dynamic database, which constitute inner state,
6. perception predicates serves as a proxy for learning facts about game environment.

## 4  Experiments and Results

In experiments *κBot* has been always compared to *E[POD]*, because *κBot* without map-specific rules and knowledge plays no different than *E[POD]*. This is caused by close translation of

*Logic Programming as Scripting Language for Bots in Computer Games – Research Overview* 7

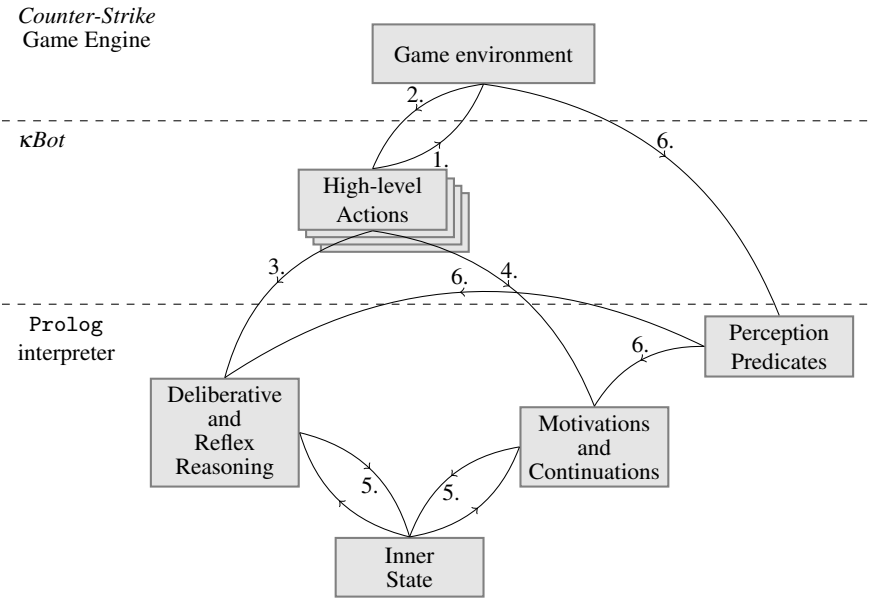

Fig. 2. Draft of bot architectural software design.

*E[POD]*'s behavioral rules, expressed by `C++` code, into `Prolog` script. This property is particularly useful, because it allows to attribute all changes in player perception of gameplay to modifications in high-level reasoning — all high-level actions, like aiming and navigating in map, were the same for the two compared bots.

### 4.1 Offline experiments

Offline experiments has been carried out, while *κBot* has been in development stage. Those kind of experiments allows to determine, if the bot is functioning effectively, *i.e.* does it not crash, does not consume too much CPU and is it able to reach its goals. Those experiments were conducted by gathering match statistics with *κBot* playing against *E[POD]*.

Those experiments were carried out on `cs_assault` map. We have tracked a number of times each team was victorious and a number of times when team fulfilled its objective. We have defined fulfilling an objective as:

- for CTs, to rescue hostages, without killing the opposing team,
- for Ts, to prevent CTs from rescuing hostages, without the opposing team.

In table 1 and table 2 we present the results of offline testing of the version of *κBot*, which was used in online experiments (section 4.2).

It could be seen that *κBot* gets better scores than *E[POD]*. Study of goal-fulfilled team victories suggests also that *κBot* is more goal-oriented than *E[POD]*.

### 4.2 Online experiments

The primary idea of online experiments was to gather feedback and metrics from people who casually play *Counter-Strike*. The metrics for players engagement were obtained through obser-vation of players game session time and receiving feedback through questionnaires. The questionnaires were about comparing gameplay quality playing with *E[POD]* bot and *κBot* bot. The volunteer played on two servers hosting `cs_assault` map: one with 7 *E[POD]* bots and one with 7 *κBot* bots in his team of choice (T or CT). The main section of questionnaire consisted of 3 comparisons: entertainment quality, realism of bot behavior and difficulty level. So far, such questionnaires has been answered for research using map-specific tactics employing virtual character negotiations. The results of questionnaires has been presented in table 3. It could be seen that this form of behavior specialization does not raise on average player gameplay satisfaction. However, the number of strong positive and negative opinions are comparable. This shows that player notice the change in bot behavior patterns. Player also feel that *κBot* behavior is more realistic and it is harder to defeat. This may explain the fact some players like it and others do not.

| CTs AI | Ts AI | CTs wins | Ts wins |
|---|---|---|---|
| *E[POD]* | *E[POD]* | 4.1 | 6.4 |
| *κBot* | *E[POD]* | 8.6 | 5.3 |
| *E[POD]* | *κBot* | 1.9 | 7.0 |
| *κBot* | *κBot* | 6.6 | 8.4 |

Table 1. *Gameplay statistics of total team victories – 10 matches average.*

| CTs AI | Ts AI | CTs wins | Ts wins |
|---|---|---|---|
| *E[POD]* | *E[POD]* | 0.4 | 1.2 |
| *κBot* | *E[POD]* | 3.6 | 0.5 |
| *E[POD]* | *κBot* | 0.2 | 2.5 |
| *κBot* | *κBot* | 1.1 | 0.1 |

Table 2. *Gameplay statistics of goal-fulfilled team victories – 10 matches average.*

| Which bot was ... | Surely *κBot* | Rather *κBot* | Hard to tell | Rather *E[POD]* | Surely *E[POD]* |
|---|---|---|---|---|---|
| ... more enjoyable? | 37.5 % | 9.375 % | 6.25 % | 18.75 % | 28.125 % |
| ... harder? | 43.75 % | 6.25 % | 15.625 % | 9.375 % | 25 % |
| ... more realistic? | 46.875 % | 9.375 % | 25 % | 6.25 % | 12.5 % |

Table 3. *The results of questionnaires in online experiments.*

The analysis of player session time shows player tend to play with *κBot* to the end of the match. When they play with *E[POD]* they tend to disconnect from server before end of match more often. This could probably, be explained by fact that player notice the difference between well-known behavior of *E[POD]* bot and new behavioral specializations of *κBot*.

### *4.3 Performance*

κ*Bot* performance was measured in order to keep resource consumption within realistic constraints for practically usable software. The presence of κ*Bot* in the online experiments is example of a server-side deployment. It differs from client-side deployment in terms of hardware and game engine functioning. Server machine usually has more CPU power than laptop or workstation. The server does not need to render 3D game environment scene, it only has to provide symbolic information about game state to its clients. Because of those difference tests were run for both settings. In general, the architecture seems to be usable: it is possible to define useful behavior without much impact on CPU. In client-side deployment `Prolog` interpreter took $\approx 1\%$ of wall time. In server-side deployment κ*Bot* caused 7% raise of CPU utilization when compared to *E[POD]*. Note that those two metrics are not comparable to each other, but each of them is appropriate to its application.

## 5 Future Works

The κ*Bot* architecture is a baseline for further extensions. We have tested a form of specialization of bot behavior using very simplified default logic. Currently we are investigating obtaining behavioral rules by computational methods.

Interpreters for some kind of logics, *e.g.* `Golog`, are implemented in `Prolog` or output a script which is executed in `Prolog` interpreter. This gives opportunities to test those logics as reasoning mechanism for κ*Bot*, without burden of integration with bot's `C++` code.

## Acknowledgements

I would like to thank my mentor prof. Jarosław Arabas for advices provided while writing this article.